\documentclass{aa}
\usepackage{epsfig,graphics,astron}
\title{Gamma-Ray Bursts and the Origin of Chondrules and Planets}
\author{B.\,McBreen \and L.\,Hanlon}
\authorrunning{McBreen and Hanlon}
\titlerunning{GRBs and the Origin of Chondrules and
Planets}
\institute{Department of Experimental Physics, University
College Dublin, Dublin 4, Ireland}
\offprints{bmcbreen@ollamh.ucd.ie}
\date{}
\thesaurus{13.07.1;  07.19.1; 07.16.1; 08.16.2; 07.13.1; 07.05.1}

\begin{document}
\maketitle

\begin{abstract}
The effect of a nearby $\gamma$-ray burst (GRB) on the
preplanetary solar nebula
 is considered.  The intense irradiation by x-rays and $\gamma$-rays would have caused
 dust balls to become molten in a matter of seconds,  cooling more slowly to
 form chondrules.  The role of iron is crucial in this process because it was the major
 absorber of x-rays between 7 keV and 30 keV.  In this
 scenario, chondrules formed at the same time across the side of the nebula
 toward the GRB source.  At least 27 Earth masses (M$\oplus$) could have been
 produced in the nebula with well mixed gas and dust of solar composition, increasing to 100 M$\oplus$ with only moderate depletion of
 nebular gases.  The chondrules combined to form meteorites and possibly the
 terrestrial planets, the cores of the giant planets and chondrules in comets.
  Assuming GRBs are linked to massive stars like supernovae, the
  probability of a GRB within 100 pc  which could form  chondrules is
  about 10$^{-3}$ and the same small probability may apply to other
  planetary systems being akin to our solar system.
  A GRB in a nearby galaxy will reveal protoplanetary disks by delayed
  transient infrared emission from the chondrule formation process.
   We suggest that a GRB was first detected about 4.5 Gyrs ago and its signature was written in stone and preserved by the chondrules in meteorites.
\end{abstract}

\section{Introduction}

Chondrules are millimetre sized, spherical to irregular shaped objects that
constitute the major component of most chondrite meteorites that
originate in the region between Mars and Jupiter and which fall to
 the Earth.  They appear to have crystallised rapidly from molten or
partially molten drops and were described \cite{sor:1877} as ``molten
drops in a fiery rain''.
The properties of the chondrules and chondrites have been exquisitely deduced from
an extensive series of experiments and two conferences
have been devoted completely to chondrules \cite{king:1983,hjs:1996}.
  The mineralogy of chondrules is dominated by olivine ((FeMg)$_{2}$SiO$_{4}$)
  and pyroxene ((FeMg)SiO$_{3}$) and there is a
wide range of compositions for all elements.  This diversity is
consistent with the melting of heterogeneous precursor solids or
dust balls.
 The age of chondrules  indicate they formed very early in the solar system.
The calcium-, aluminium-rich inclusions (CAIs) are refractory inclusions in
carbonaceous and ordinary chondrites that predate the chondrules by several
million years and are the oldest known solid materials produced in the nebula
\cite{swin:1996}.  The first 10$^{7}$ years in the complicated development of
the solar system has been comprehensively covered in an attempt to understand
the CAI to chondrule time interval of several million years \cite{cam3:1995}.

The presence of volatile elements in the chondrules indicate that
the high temperature melting period
lasted for a matter of seconds to minutes.  Experiments based on chemical
and textural compositions of chondrules suggest cooling rates that were
much slower than radiative cooling of isolated chondrules and imply
they were made in some large quantity in relatively opaque
nebular domains \cite{yuh:1998}.  Volatile elements
such as alkalis and sulphur occur in chondrule interiors as primary
constituents and indicate that some chondrule precursor materials must
have reacted with cool nebula gases at ambient temperatures less than 650 K.

The heat source that melted the chondrules remains uncertain and a critical
summary of the heating mechanisms was given by Boss \cite*{bos2:1996}.  These
methods include giant lightning flashes \cite{hmg:1995} and shock wave heating
of the precursor materials \cite{wo:1988}. All heat sources proposed to form
the chondrules are local to the solar nebula.  We propose that the chondrules
were flash heated to melting point by a nearby GRB when the precursor
materials efficiently absorbed x-rays and low energy $\gamma$-rays.  The
distance to the source was about 300 light years (or 100 pc) for a GRB output
of 10$^{53}$ ergs and was estimated using the minimum value of \(2 \times
10^{10}\) erg g$^{-1}$ required to heat and melt the precursor grains
\cite{grn:1988,was:1993}.  The role of nearby supernovae that preceded the
formation of the solar system have been considered \cite{chmc:1995} along with
the serious consequences for life on Earth of nearby supernovae
\cite{rud:1974,cmcs:1977} and GRBs \cite{thor:1995}.  The consequences of a
nearby GRB on the early solar nebula have not been considered elsewhere.

The properties of GRBs relevent to chondrule formation are
presented in section 2.  The absorption of x-rays and
$\gamma$-rays by gas and dust in the solar nebula is considered in
section 3. The effects of sudden chondrule production on the
formation of the planets are presented in section 4.  The
probability of a GRB producing chondrules is considered in section
5.

\begin{figure}
\resizebox{\columnwidth}{!}{\includegraphics{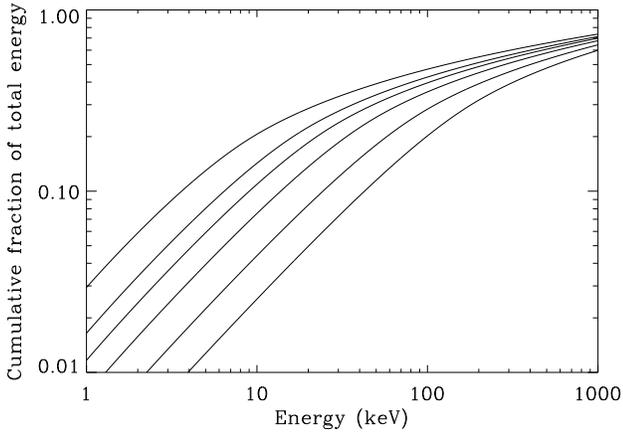}} \caption{The
cumulative fraction of GRB energy as a function of photon energy for assumed
spectral parameters $\alpha$ = -1, $\beta$ = -2, $E_{o}$ = 5, 10, 15, 25, 50
and 100 keV and redshift of 0.8.}
\end{figure}

\section{$\gamma$-Ray Bursts (GRBs) as a Heat Source}

Since their discovery thirty years ago the properties of GRBs have been
determined by an outstanding series of experiments that were deployed on more
than twenty spacecraft at distances up to several astronomical units (AU) from
the Earth and during one period 11 spacecraft were used to study the same
GRBs. The properties of GRBs have recently been reviewed
\cite{fm:1995,pir:1999} and are now the subject of intense research.  GRBs are
extragalactic in origin and release collossal amounts of energy, \(10^{52}\)
to \(4 \times 10^{54}\) ergs assuming isotropic emission, for the GRBs with
known redshift.  The BeppoSAX satellite discovered x-ray afterglow that
decayed with time t in the range t$^{-1.1}$ to t$^{-1.5}$
\cite{cfh:1997,piro:1998,nic:1998}. Simultaneous optical emission was detected
from the spectacular GRB 990123 at the level of about \(10^{-5}\) of the
energy in $\gamma$-rays \cite{abb:1999}, but this emission is too weak to
influence chondrule formation.

The progenitors of GRBs are not known but merging neutron stars have been
suggested. One GRB is known to have occurred in a luminous infrared galaxy
\cite{han:1999}, and some GRBs are close to star forming regions suggesting a
connection with massive stars \cite{bd:1998}. Models of `failed supernova' and
`hypernova' have been proposed \cite{woo:1993,pac:1998} in which the inner
core of a massive rotating star collapses to a black hole while the outer core
forms a massive disk or torus that somehow generates a relativistic fireball
and GRB. In these models GRBs represent the violent end to massive stars.

GRBs have a lognormal bimodal distribution of durations that peak at about 0.3
s and 30 s respectively.  The wide range of pulse shapes with complex time
profiles have been described using lognormal distributions and GRBs have been
called cosmic lightning because of their statistical similiarities with
terrestrial lightning \cite{mhl:1994,hur:1998,ss:1996}. The photon spectra are
well described by a power-law with a low energy slope $\alpha$, a break energy
E$_{o}$ and a high energy power law with slope $\beta$. The functional form,
for low energies, is given by \cite{bmf:1993} :
\[N(E) = A E^{\alpha}e^{-E/E_{o}} \, \, \, {\rm and} \,
\, \, N(E) = B E^{\beta}, \hspace{.3cm} \alpha > \beta\] The value
of $E_{o}$ ranges from 2 keV to over 1 MeV and the indices
$\alpha$ and $\beta$ are typically -1 and -2 respectively. There
is an excess of $\approx$ 10 keV x-rays above this functional form
in about 15\% of GRBs \cite{pbp:1996}.

The spectral energy distributions of a sample of GRBs have been extrapolated
and integrated from 0.1 keV to 10 MeV using a sample of values for $E_{o}$
compatible with BATSE \cite{bmf:1993} and Ginga results \cite{sfm:1998}. The
cumulative fraction of the total energy in GRBs is given in Figure 1 where a
redshift correction of z = 0.8 has been applied to all the spectra. There is
sufficient energy to melt solar nebula grains provided absorption of the
x-rays and $\gamma$-rays is reasonably efficient.

\section{Attenuation of X-Rays and $\gamma$-Rays}

The absorption of the GRB energy by the gas and dust in the nebula
would have occurred through the processes of photoelectric
absorption and Compton scattering. The combined cross-sections due
to these processes for the elements from H to Fe are plotted in
Figure 2. Solar abundance values were adopted for the nebula
\cite{ag:1989} and the photoelectric and Compton cross-sections
for the elements have been used
 \cite{vei:1973} with the exception of molecular hydrogen (H$_{2}$)
 where a value of 1.25
times the elemental photoelectric cross-section was adopted
\cite{mmcc:1983}. The photoelectric effect absorbs the photon
completely but in Compton scattering only a fraction of the energy
is removed per scattering and this fraction varies from 0.14 at
100 keV to 0.45 at 1 MeV. The product of the Compton cross-section
by the fraction of the energy absorbed in the first scattering was
used for the Compton cross-section and many scatterings may occur
before the degraded photon is finally absorbed by the
photoelectric effect. Decreasing the abundance of H and He by a
factor of 10 relative to the solar value significantly reduces the
cross-section below 1 keV and above 20 keV (Figure 2b).  The
cross-section of Fe (Figure 2d) is dominant between the K edge at
7.1 keV and 30 keV but the upper bound of 30 keV extends to above
50 keV for low abundance of H and He. Fe makes the major
contribution to the absorption by the dust and is the key to
 chondrule formation.

\begin{figure}
\resizebox{\columnwidth}{!}{\includegraphics{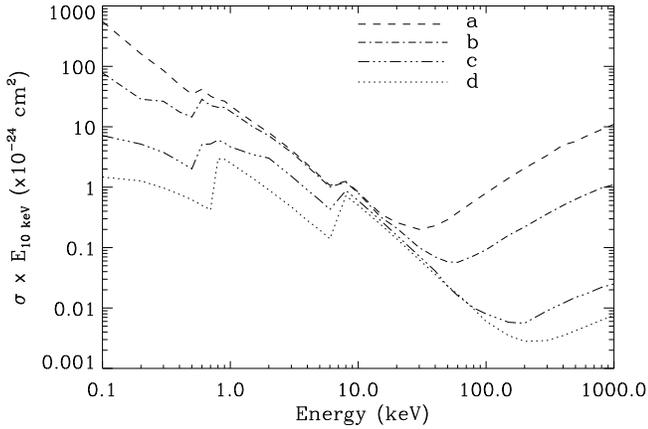}} \caption{The
combined photoelectric and Compton cross sections relative to H as a function
of energy, scaled by E/10 keV for clarity of presentation. (a)  The elements H
to Fe that are at least as abundant as Fe (H, He, C, N, O, Ne, Mg, Si, Fe) (b)
the same elements as in (a) with solar abundance of H and He reduced by 10 (c)
The precursor dust combination Fe$_{0.9}$SiMg$_{1.1}$O$_{4}$ and (d) the
element Fe.}
\end{figure}

The elements O, Si, Mg and Fe dominate the composition of the
chondrules but the composition of the precursor grains has been
the subject of much study and speculation \cite{hew1:1997}. A
number of chondrule classification systems have been adopted and
early approaches depended mainly on either bulk composition or
texture or both. McSween \cite*{mcs:1977} recognised two main
types, type I or FeO poor and type II or FeO rich. However in
comparison with chondrules, the fine-ground matrix in chondritic
meteorites is more FeO rich than even type II chondrules and it
has been proposed that this matrix may be close to the composition
of the chondrule precursor \cite{hlpw:1996}. In this scenario, the
precursors of type I chondrules were enriched in Fe and more
efficiently heated by x-ray and $\gamma$-ray absorption, resulting
in the loss of Fe and other volatile elements that ended in the
enriched matrix.  Type II chondrules appear not to have lost
significant amounts of volatiles in the melting process.  The
composition of the precursor grains may be resolved by new x-ray
and $\gamma$-ray heating experiments. A composition consisting of
solar abundance of oxides of Fe, Si and Mg or
Fe$_{0.9}$SiMg$_{1.1}$O$_{4}$ was assumed for the precursor grains
and the product of the cross section of this combination by
abundance relative to H is plotted in Figure 2c. The x-ray and
$\gamma$-ray absorption efficiencies of different thicknesses of
precursor grains, assuming a density of one, are given in Figure 3
and grains in the range
 10 $\mu$m to 1 cm are very efficient absorbers in the region
 where dust dominates the absorption (Figure 2).  This range agrees
 quite well with the measured Weibull and lognormal distributions of
  chondrule sizes \cite{mh:1980,rk:1984}.
The deficiency of small grains is caused by low absorption efficiency and substantial radiation
losses from grains with large surface to volume ratios.

The thickness of the dust layer converted to chondrules depends on the GRB
spectrum which must have significant emission below  30 keV where dust
absorption dominates (Figure 2) and also on the mixture and distribution of
gas and dust in the nebula.  For a GRB with 10$^{53}$ ergs and an assumed
spectrum $\alpha$ = -1, $\beta$ = - 2 and $E_{o}$ = 15 keV (Figure 1), the
fraction of GRB energy photoelectrically absorbed by the dust is 20\%,
increasing to 27\% for a factor 10 reduction in H and He. In the simplified
case of solar abundance and a uniform mix of gas and dust,  the thickness of
the chondrule layer created is 0.18 g cm$^{-2}$ corresponding to one optical
depth for 30 keV x-rays. The layer thickness increases to about 0.8 g
cm$^{-2}$ and 2.0 g cm$^{-2}$ for optical depths to 40 keV and 55 keV x-rays
with H and He abundances reduced by factors of 3 and 10 respectively. The
thickness of the chondrule layer is therefore controlled by the degree of gas
depletion from the nebula.  The minimum GRB fluence required to produce
chondrule layers of 0.18, 0.8 and 2.0 g cm$^{-2}$ is 1.8 $\times$ 10$^{10}$,
7.0 $\times$ 10$^{10}$ and 1.5 $\times$ 10$^{11}$ ergs cm$^{-2}$, adopting
20\%, 23\% and 27\% absorption by the chondrule precursors and 2 $\times$
10$^{10}$ erg g$^{-1}$ for heating
 and melting.  A fluence of 10$^{11}$ erg cm$^{-2}$ implies a distance of
 about
 100 pc to the source for an output of 10$^{53}$ ergs radiated isotropically.
   The GRB would also form a layer of chondrules over a large area
   (10$^{3}$ - 10$^{4}$ pc$^{2}$)
   in a nearby molecular cloud
   provided large precursor grains had already formed \cite{wr:1994}.  The
   process of chondrule amalgamation might be sufficient to
   trigger star formation over this region. In this case
   chondrule formation precedes cloud collapse and star formation.  The
   existence of pre-solar grains in meteorites is well established
   \cite{zin:1996} but there is no evidence for pre-solar chondrules.

The chondrules cooled at a much slower rate than if they were isolated
\cite{hew1:1997}.  They may have been warmed by a fading source or by forming
 a thermal blanket or a combination of both effects.  BeppoSAX discovered
 x-ray afterglow from GRB sources and the limited measurements show
 considerable variability between the various GRBs \cite{cfh:1997,piro:1998}.  The
 afterglow typically decreases by at least a factor of 20 in 10$^{3}$ s which yields more than a factor of two drop in temperature.
This decrease of 3800 K hr$^{-1}$ is too rapid to account for the
 chemical and textural properties of chondrules \cite{yuh:1998}.  The
 chondrules on the far side of the layer from the GRB source  cool even more
 rapidly because of
 shielding by foreground chondrules and spectral softening of the afterglow.
The optical depth of the chondrules to their infrared radiation at
a peak of about 1.5~$\mu$m is about 0.25 g cm$^{-2}$ assuming all
the chondrules have size 0.1 cm \cite{hoh:1991,wo:1988}.  The
x-ray and infrared optical depths are comparable and the cooling
rate was further reduced by this thermal blanket.

There are several indicators that the dust was concentrated and/or gas
depleted in the nebula when chondrules formed.  These include: (1) the
 increased rate of collisions between plastic and molten chondrules to
 form adhering pairs \cite{was:1993}, (2) the seeding of melted chondrules
 with dust grains \cite{ch:1995}, (3)  the O/H ratio well above the solar
 value \cite{fp:1985}, and (4) the improved absorption efficiency of x-rays
 and $\gamma$-rays by the precursor dust balls.  The rims on chondrules indicate time
 spent in dusty regions.

\section{The Solar Nebula and Chondrules}

The formation of the Sun and planets has been the subject of
extensive study and is now particularly important because of the
recent detections of Jupiter like planets around solar type stars
that are inferred to be giant planets
\cite{mq:1995,bs:1996,LLD:1998}. The main model for planetary
formation in the solar system is that planets are the end result
of a bottom-up assembly process beginning with the accumulation of
interstellar grains into millimeter and centimeter sized objects
that form in the disk and eventually settle to the midplane where
they are brought together to form kilometer sized objects
\cite{miz:1980,bl1:1995}.  These so called planetesimals proceed
through a runaway accretion process to form bodies of lunar size
that are eventually accumulated to form the terrestrial planets.
In the case of the giant and icy outer planets the process is
different \cite{phb:1996}.  A planetary core accretes until a
critical mass of about 10 M$\oplus$ is reached.  At this point the
growing core is unable to sustain an equilibrium atmosphere and a
very rapid accretion of nebular gas occurs that gives a giant
planet. The total mass of the nebula between 0.35 AU and 36 AU is
0.01 M$\odot$ and  is considered the minimum necessary to form the
present planets. There are perceived difficulties with this model
with regard to timescales because the predominantly H and He
composition of Jupiter and Saturn predates the dispersal of the
solar nebula \cite{cam2:1978}.

\begin{figure}
\resizebox{\columnwidth}{!}{\includegraphics{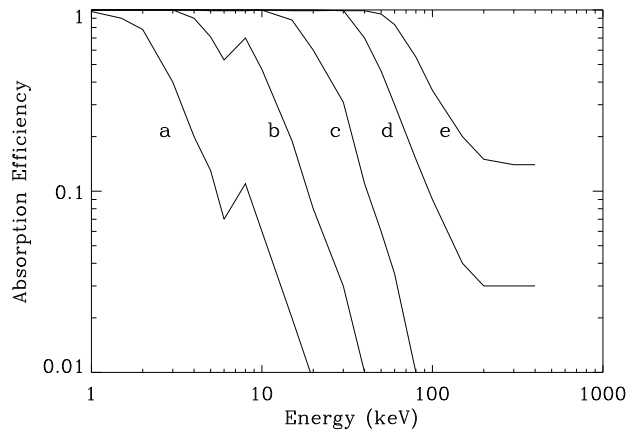}} \caption{The
absorption efficiency of different thickness of Fe$_{0.9}$SiMg$_{1.1}$O$_{4}$
as a function of energy:  (a) 10 $\mu$m, (b) 100 $\mu$m, (c) 1000 $\mu$m, and
(d) 1 cm, (e) 5 cm.}
\end{figure}

A number of models have been proposed for distribution of surface
densities of dust materials and gases in a preplanetary solar
nebula, which is in thermal and gravitational equilibrium.
According to  Cameron \cite*{cam3:1995} the nebula surface density
as a function of radial distance from the Sun was $\Sigma =
\Sigma_{0}$ r$^{-1.0}$  g  cm$^{-2}$ with a normalising value of
\(\Sigma_{0} = 4.25 \times 10^{3}\) at r = 1 AU.  The abundance by
mass of rocky and metallic materials was 0.0043 and 0.0137 for icy
materials (H$_{2}$O, CH$_{4}$ and NH$_{3}$), that occur beyond
about 3 AU.

There are many complicated processes that occur in the evolution of the solar
nebula from essentially interstellar grains to planetesimals and planets (e.g.
\cite{cam3:1995}).  Particle radii vary over 11 orders of magnitude from
micron sized interstellar particles to planetesimals.  It seems likely that
planetesimal formation involved at least an early stage of collisional
sticking and coagulation of particles.  This process depended on poorly
constrained properties of the nebula such as turbulence and of the particles
themselves such as stickiness  and composition.  The process of particle
accumulation and coagulation is strongly influenced by turbulence which keeps
the particles in constant random motion with respect to the gas and many
simulations on particle coagulation have been carried out for silicate and icy
grains \cite{cam3:1995,weid:1980,wc:1993}.  In the GRB-chondrule scenario, the
chondrules reflect the size distribution of the precursor grains.  It is
interesting that they have an approximate lognormal distribution.  This
distribution is generated by particle grinding and collisions and imply these
conditions prevailed in the nebula at the time of the GRB \cite{ab:1957}.
Historically the lognormal distribution was developed to account for the size
distribution of grains of sand.  Turbulence is known to produce structure on
many scales and hence the particle distribution in the nebula was probably
quite clumpy.  To make progress we assume the particles and gas were well
mixed and of solar composition.
 The GRB should have produced a layer of
0.18 g cm$^{-2}$ of chondrules or 27 M$\oplus$ out to r = 36 AU and 100
M$\oplus$ if H and He were depleted uniformly by a factor of 3.   In the
latter case, three percent and ten percent of the dust is converted to
chondrules at r = 1 AU and r = 10 AU respectively.  The gas may not have been
depleted uniformly across the nebula at the time of the GRB. The compositions
of Jupiter and Saturn reveal that they formed before the nebula was depleted
of H and He by more than about a factor of 5.  The terrestrial planets and
giant icy outer planets, Uranus and Neptune, completed their formation after
gas depletion from the nebula.

The gas depletion from the inner nebula probably occurred over about 10$^{7}$ years by strong solar winds during the FU
Orionis and T-Tauri phase of the Sun and from the outer nebula beyond about 9 AU by
photo evaporation \cite{sjh:1993}.
In this scenario the chondrules formed almost simultaneously across the side of the
nebula towards the GRB and impacted on the early history of aggregates in the nebula.
The chondrules were combined and compacted to form meteorites
and this same process should have operated throughout the nebula and led
to the formation of the terrestrial planets and possibly the cores of the
 giant planets.
The minimum of 27 M$\oplus$ is close to the amount estimated
(30-40 M$\oplus$) to form the cores of the four giant planets
assuming that chondrules constituted most of the mass.  However
for chondrules to form the terrestrial planets, then a layer of at
least 2 g cm$^{-2}$ was produced out to a few AU, implying
significant gas dispersal from the inner nebula by about a factor
of 10 at the time of the GRB.  The composition of the Earth is not
inconsistent with a complete chondrule origin because the inner
solar system experienced higher ambient temperatures than the
asteroid belt and would have been more efficiently cleaned of gas
containing volatile bearing dust  \cite{hh:1996}.

There is a  major change in the aerodynamic properties on melting of the
fluffy precursor material to form chondrules with stopping times increasing by
a  factor of about 100.  However, there seems to be little tendency for them
to settle to the midplane because vertical settling requires very low
intensities of turbulence \cite{weid:1980,vm:1991,cam3:1995}. Somehow the
chondrules were size sorted, probably by an aerodynamic process, and
concentrated by a large factor to form planetesimals and meteorites of size of
order 100 km.  A turbulent concentration of particles has been shown to be
size-selective and effective for particles with the chondrule size
distribution \cite{cd:1996} where a uniform volume distribution of particles
tend to vacate the eddies and concentrate in stagnant zones with concentration
factors of order 10$^{6}$. The subsequent evolution of the dense
concentrations has not been modelled in detail but it is probable that they
descended to the midplane and commenced and continued the accretion process of
forming planetesimals that include the meteorite parent bodies. Subsequent
collisions must merge the planetesimals of size about 100 km to form the
terrestrial planets and cores of giant planets. The formation of the cores of
the giant planets could have proceeded at a much faster rate than the
terrestrial planets because the gas was cooler, less dense and less turbulent
and also the chondrules were probably
 coated with ice and mixed with ice
particles that should have significantly improved the sticking
co-efficients \cite{cam3:1995}.  These effects should have speeded up the formation of planetesimals in the outer nebula.

The composition of comets is approximately solar and the GRB should have formed chondrules in the region where comets formed.
The favoured region is near the Uranus-Neptune zone where
perturbations by the proto-neptunian group could move the young comets
out to the Oort cloud \cite{whi:1989,mum:1993}.  Other models advocate cometary formation
further out in the solar system.  There are only two ways in which comets
can avoid having chondrules:  (1) they formed before the GRB, or (2) there
were no iron rich dust balls sufficiently large to be melted to form chondrules.
The presence or absence of chondrules in comets will yield valuable
clues to the cometary and chondrule formation processes.  The high precision results that will come from outstanding and ambitious rendezvous and sample return missions will greatly improve our understanding of conditions in the comet forming regions of the solar system.

\section{Probability of a nearby GRB}

The frequency of nearby supernovae, and hence GRBs assuming they are linked to
massive star formation like supernovae, depend on where the solar system was
located within the galaxy when it formed. The highest rate of type II
supernovae occurs in the two principal spiral arms of the galaxy. The
molecular cloud was
 compressed entering the spiral arm to a condition for star
formation and this interaction resulted in a new star cluster that traversed
the spiral arm.  Massive stars in the cluster evolve rapidly over \(10^{7}\)
years terminating in type II supernovae. The width of this supernova zone is
about 1 kpc because the stars move at about 100 km s$^{-1}$ for \(10^{7}\)
years \cite{cmcs:1977}.  It is likely that a nearby supernova caused the
collapse of the presolar cloud and also seeded the nebula with the radioactive
$^{26}$Al needed to explain the \(^{26}\)Mg in CAIs \cite{chmc:1995}.  The
number of supernovae along the spiral arm, within the 1 kpc zone and over a
period of \(10^{7}\) years, has been estimated at 250 supernovae per 100 pc
\cite{cmcs:1977}.

BATSE observes on average about one GRB per day.  This corresponds to one
burst per million years per galaxy assuming that the rate of GRBs does not
change with cosmological time \cite{fm:1995}.  The average rate changes if
allowance is made for beaming or a cosmic evolution of the rate of GRBs. The
observations that GRB host galaxies are star forming systems
\cite{hf:1998,ftm:1999,bd:1998} indicates that the rate of GRBs may follow the
star formation rate \cite{wbb:1998,tot:1999}.  In this case GRBs are further
away and occur at a lower rate and have significantly greater energy output.
 At present there is no agreement on the nature of the progenitors of the GRB
explosion although neutron star mergers are a promising candidate
\cite{eich:1989,pir:1999}.  The list also include failed supernovae
\cite{woo:1993}, white dwarf collapse \cite{usov:1992} and hypernovae
\cite{pac:1998}.  All these models are consistent with the possibility that
GRBs are associated with star forming regions.  The lifetime of massive stars
is quite short and that of a neutron star binary could be sufficiently short
to be close to a star forming region.

There is considerable uncertainty in the cosmological rate of GRBs
\cite{cen:1998,kth:1998,che:1999} and a rate of one GRB per galaxy per
10$^{7}$ years is adopted which is about 10$^{5}$ times less than the
supernova rate \cite{pac:1998}. It is also assumed that GRBs are linked to
massive stars and the explosion occurs in the supernova zone of the spiral
arm. There is a probability of about 0.001 of a GRB occuring within $\pm$ 100
pc of the solar nebula assuming the length of the spiral arms is about 40 kpc
and the thickness of the spiral arm perpendicular to the plane is less than
100 pc. The probability will be smaller by many orders of magnitude if GRB
explosions occur at random locations throughout the galaxy.  There is evidence
such as paired and rimmed chondrules that some of them were melted on more
than one occasion \cite{hew1:1997,was:1993}.  The probability of two GRBs
impacting on the solar nebula with sufficient energy to melt chondrules is
\(10^{-6}\).  The heat source that led to CAI formation is uncertain but it
was much more intense and lasted for a longer period than chondrules because
most of the refractory dust was evaporated in the process \cite{wo:1988}.  A
GRB could have been the heat source but it is very improbable because it must
have been within 10 pc to provide the required energy.

If this GRB-chondrule scenario is correct,
 then only about one planetary system in 1000 should have
evolved like the solar system and should preserve evidence for chondrule
formation. The solar nebula existed as a detector of intense flashes of
radiation for millions of years but recent satellite observations cover less
than forty years and have discovered the GRBs and soft $\gamma$-ray repeaters
(SGRs). There could be other rare transient sources yet to be discovered that
influenced the formation of chondrules. In this context the role of the SGRs
might have been important \cite{kds:1993}.  There are four known SGRs that are
associated with supernova remnants and which have high velocities relative to
the nebula.  Two of the SGRs have generated intense transients, \(5 \times
10^{44}\) ergs and \(2 \times 10^{43}\) ergs, but these transients are too
feeble by about a factor of \(10^{6}\) to influence chondrule formation
\cite{hcm:1999}. However the number of SGR sources within the galaxy is very
uncertain \cite{hmr:1994,mcbh:1998,hk:1998} and SGRs may generate much more
powerful outbursts shortly after their formation.  The recent detection
\cite{gvv:1998} of a weak GRB, about \(10^{48}\) ergs, from a type Ib/c
supernova suggests that different mechanisms may give rise to a new class of
dim supernova-related GRBs.

A GRB in a nearby galaxy ($<$ 100 Mpc) could be used to reveal
protoplanetary disks because of the transient infrared emission
from chondrule formation.  In K band, the transient source would
be at the $\mu$Jy level and good angular resolution is required to
separate the transient emission from the galactic background.  The
transient sources could occur over a period of hundreds of years
after the GRB, assuming isotropic GRB emission.

\section{Conclusions}

It has been shown that a nearby GRB could have melted pre-existing dust balls
in the pre-planetary solar nebula and produced chondrules across the nebula at
the same time.  The probability of a GRB occurring within $\approx$100 pc is
about \(10^{-3}\). Sufficient chondrules could have been  produced by this
mechanism to account for the meteorites and cores of the giant planets and
enough chondrules to completely account for the Earth and the inner planets
provided the gas in the inner solar system was depleted by a factor 10. The
probability of other planetary systems being similar to the solar system is
about \(10^{-3}\). A GRB in a nearby galaxy could be used to find
protoplanetary disks by detecting the transient emission from chondrule
formation.  Chondrule layers could also be formed in molecular clouds that are
near a GRB and have large precursor grains.

\begin{acknowledgement}
We are pleased to thank FORBAIRT for financial support and C. Handley for her
help in the preparation of this manuscript.
\end{acknowledgement}
\bibliography{kates2}
\bibliographystyle{aanda}

\end{document}